\documentclass[aps,pra,showpacs,amsmath,amssymb,floats,twocolumn]{revtex4}
\usepackage{graphics,dcolumn,bm}
\usepackage{color,booktabs}
\usepackage[hypertex]{hyperref}

\begin{document}
\title{Ionization heating in rare-gas clusters under intense XUV laser pulses
}
\author{Mathias Arbeiter}
\author{Thomas Fennel}\email{thomas.fennel@uni-rostock.de}
\affiliation{Institute of Physics, University of Rostock, 18051 Rostock, Germany}
\date{\today}

\begin{abstract}
The interaction of intense extreme ultraviolet (XUV) laser pulses \mbox{($\lambda=32\rm\,nm$, $I=10^{11-14}$\,W/cm$^2$)} with
small rare-gas clusters (Ar$_{147}$) is studied by quasi-classical molecular dynamics simulations.
Our analysis supports a very general picture of the charging and heating dynamics in finite samples under short-wavelength radiation that is of relevance for several applications of free-electron lasers.
First, up to a certain photon flux, ionization proceeds as a series of direct photoemission events producing a jellium-like cluster potential and a characteristic plateau in the photoelectron spectrum as observed in [Bostedt {\it et al.}, Phys. Rev. Lett. {\bf 100}, 013401 (2008)]. Second, beyond the onset of photoelectron trapping, nanoplasma formation leads to evaporative electron emission with a characteristic thermal tail in the electron spectrum. A detailed analysis of this transition is presented.
Third, in contrast to the behavior in the infrared or low vacuum ultraviolet range, the nanoplasma energy capture proceeds via {\it ionization heating}, i.e., inner photoionization of localized electrons, whereas collisional heating of conduction electrons is negligible up to high laser intensities. A direct consequence of the latter is a surprising evolution of the mean energy of emitted electrons as function of laser intensity.
\end{abstract}

\pacs{52.50.Jm, 36.40.Wa,41.60.Cr}

\maketitle

\section{Introduction} \label{sec_introduction}
The rapid advances in producing ultrashort and intense short-wavelength radiation keep stimulating today's laser-matter research. Whereas sub-fs vacuum ultraviolet (VUV) pulses generated from ultrashort near infrared (NIR) laser fields have become a key tool in attosecond physics \cite{KraRMP09}, coherent and highly intense pulses from free electron lasers (FEL) are currently pushing the frontiers of intense laser-matter science from the VUV up to the x-ray domain~\cite{AyvEPJD06,AckNPhotonics07,LCLS,XFEL,RikenXFEL}. Promising examples of novel applications that are expected to become feasible with upcoming x-ray FEL's are diffractive or holographic imaging of finite samples with unprecedented spatial and temporal resolution~\cite{NeuN00,ChaN07,ChaNP06}.
A key fundamental issue being of relevance for many applications of short-wavelength FEL light is how the nature and dynamics of intense laser-matter interactions develop as function of wavelength, e.g., from the infrared towards the x-ray domain.

Promising prototype systems for corresponding exploratory studies on finite targets are atomic clusters, as great experience in their experimental and theoretical investigation is available~\cite{Krainov2002,SaaJPB06,FenRMP10}. Their efficient absorption in the well-studied NIR range ($\lambda \sim 800\,$nm) results in energetic reaction products, such as, highly charged ions~\cite{LezPRL98,SnyPRL96}, fast electrons~\cite{SprPRA03,KumPRA03}, x-rays~\cite{McPN94, PriPRA08}, or even neutrons~\cite{DitN99}. In this domain the response is governed by a strongly confined and short-lived nanoplasma, whose formation is typically triggered by field ionization. Particularly strong absorption occurs for resonant collective excitation of the nanoplasma electrons. Intense laser-cluster experiments at shorter wavelengths, i.e., in the VUV (100\,nm), became feasible by turn of the millennium at the DESY free electron laser (FLASH) and have shown unexpectedly strong heating and highly charged ion production~\cite{WabN02}, though collective plasma heating effects become negligible due to low quiver amplitudes of electrons. In particular, strong polarization fields and resonance enhancement as typical for the NIR disappear~\cite{BauJPB04}. A further fundamental difference of the VUV domain is that vertical photoionization becomes the trigger ionization process for the nanoplasma formation. As electronic screening and local fields from neighboring ions in the cluster lower the atomic ionization barriers, cluster constituents can be charged much stronger than isolated atoms with the same photon energy via photoionization~\cite{SiePRL04,GeoPRA07}. Taking this effect into account, the observed cluster heating could be explained by inverse Bremsstrahlung (IBS) in the resulting dense nanoplasma~\cite{SaaJPB06}. Additional aspects that have been put forward to explain strong absorption in the VUV include the effects of realistic ion potentials for describing IBS~\cite{SanPRL03} and enhanced heating through many-body collisions in a transient strongly coupled nanoplasma~\cite{JunJPB05}. A general conclusion being in line with the observed thermal electron spectra~\cite{LaaPRL05} is that laser heating of quasifree electrons via IBS is a leading absorption process in the \mbox{$\lambda\approx 100\,$nm} regime. Simulations on Ar$_{\rm N}$ support that IBS remains substantial even with $\lambda=60\,$nm~\cite{GeoPRA07}.

Evidence for a transition to a new response regime have been found in experiments in the extreme ultraviolet range (XUV)~\cite{BosPRL08}, where photoelectron spectra from rare-gas clusters \mbox{(Ar$_{\rm N}, \langle{\rm N}\rangle=80$)} under 30\,fs FEL pulses at \mbox{$\lambda=32\,$}nm showed pronounced nonthermal, plateau-like signatures. Up to laser intensities of about $10^{13}{\rm W/cm^2}$ the experimental electron spectra can be well explained with a multistep ionization model that excludes nanoplasma effects and assumes sequential atomic photoionization in the developing cluster Coulomb potential. An additional low energy tail emerging for higher intensities could only partly be explained with the multistep model, suggesting a dynamical mixture of direct photoemission and thermal electron evaporation.

The purpose of the current study is to examine the intensity-dependent ionization and heating processes at \mbox{$\lambda=32\,$}nm \mbox{($\hbar\omega=$38\,eV)} in more detail. A quasi-classical molecular dynamics approach including plasma effects like atomic ionization threshold lowering through local plasma fields is applied to the model system Ar$_{\rm 147}$. The simulation results indicate that the energy capture of the nanoplasma at high fluences is dominated by a new mechanism we term {\it ionization heating}. Therein, the absorption is no longer mediated by laser heating of nanoplasma electrons, but by inner photoionization of remaining localized electrons. A clear marker for this heating mechanism in the electron emission spectra is discussed.

The remainder of the text is organized as follows. Methodical details of the simulations are subject of Sec.~\ref{sec_methods}. Section.~\ref{sec_results} proceeds with a presentation of numerical results and discusses characteristic phases of the interaction process, the intensity dependence of electron energy spectra, and signatures from ionization heating. Conclusions are drawn in Sec.~\ref{sec_conclusions}.

\section{Methods} \label{sec_methods}
As a full quantum mechanical description of clusters in strong laser fields is currently out of reach, quasi-classical approaches are widely used to model the interaction process, see, e.g., Refs.~\cite{SaaJPB06,FenRMP10} and references therein. The common strategy is to describe atomic ionization processes quantum mechanically via appropriate rates, whereas the dynamics of the resulting ions and electrons is treated classically. The classical propagation of electrons is justified for sufficiently strong excitations and has been applied with great success to laser-cluster interactions from the NIR up to the x-ray domain. We thus follow this general strategy in our simulation code.

Starting point of a simulation run is the initialization of an Ar$_{147}$ cluster in relaxed icosahedral structure. For the annealing a binary Lennard-Jones potential has been used \cite{Ogilvie1992}. At this stage only neutral atoms appear explicitly in the code. As optical field ionization is negligible in the XUV range, only atomic photoionization is included as inner ionization process. We assume sequential one-photon absorption to be the dominant channel and neglect multi-photon ionization. Ionization events are sampled statistically in each timestep $\Delta t$ using a Monte-Carlo scheme with the probability for the $k$-th atom/ion
\begin{equation}
\mbox{$p_k=1-\exp(1-W_k\Delta t)$},
\end{equation}
where \mbox{$W_k=\sum_{\alpha}\sigma_k^{\alpha} \,I(t)/{\hbar\omega}$} gives its actual ionization rate for the instantaneous laser intensity $I(t)=I_{0}\,f\,^2(t)$ summed over the relevant atomic levels $\alpha$ (3s and 3p levels are included here) with the corresponding photoionization cross sections $\sigma_k^{\alpha}$. Here $I_{0}$ and \mbox{$f(t)=\exp(-2\ln2 \,t^2/\tau^2)$} denote the peak intensity and the normalized temporal field envelope of a gaussian laser pulse with pulse duration $\tau$ (FWHM), respectively.

The photoionization cross sections are taken from free atomic Ar, see Tab.\,\ref{tab_argon}, for which ${\rm Ar}^{2+}$ is the maximum charge state that can be attained via one-photon photoionization with \mbox{$\hbar\omega=38$\,eV}.
\begin{table}
\renewcommand{\arraystretch}{1.15}
\begin{tabular}{l | c  c  c  c  c c}
   & ${\rm occ}^{3s}$  & ${\rm occ}^{3p}$ & $I_p^{3s}$ [eV] & $\sigma^{\rm 3s}$\,[Mb] & $I_p^{3p}$ & $\sigma^{3p}$\,[Mb]   \\
\hline
Ar        & 2  & 6  & $29.3$            &      $0.2^a$                    & $15.76^*$          &  $5.0^a$    \\
Ar$^{+}$  & 2  & 5  & $43.67$           &      -                   & $27.63^*$          &  $2.6^b$    \\
Ar$^{2+}$ & 2  & 4  & $58.09$           &      -                   & $42.54$         &   -  \\
Ar$^{3+}$ & 2  & 3  & $73.60$           &      -                   & $57.60$         &   -  \\
Ar$^{4+}$ & 2  & 2  & $90.07$           &      -                   & $74.96$         &   -  \\
Ar$^{5+}$ & 2  & 1  & $107.4$          &      -                   & $90.94$        &   -
\end{tabular}
\caption{\label{tab_argon}Level occupation ($\rm occ$), ionization energies ($I_p$), and photoionization cross sections ($\sigma$) at \mbox{$\hbar\omega=38\,\rm eV$} of atomic Ar as function of charge state. Electron removal from the 3s or the 3p level is considered (as indicated). Values with superscripts have been taken from $^*$NIST, $^a$\cite{BeckerShirley}, and $^b$\cite{Covington2001}. Remaining ionization potentials are calculated with an atomic all-electron dirac-LDA code \cite{Ank1996}.}
\end{table}
In a cluster, however, further ionization is possible because of the presence of the local plasma environment~\cite{SiePRL04,GeoPRA07}. Considering an atomic ion in the charged cluster, neighboring ions and electronic screening by quasifree cluster electrons lead to an effective ionization threshold $I_p^*=I_p-\Delta_{\rm env}$ that is lowered with respect to the pure atomic value by the environmental shift $\Delta_{\rm env}$. Here $I_p^*$ specifies the minimum energy required to lift the electron into the quasi-continuum of the cluster. The shift $\Delta_{\rm env}$ is evaluated directly from the plasma field in the MD simulation using the scheme developed in Ref.~\cite{FenPRL07b}. Therefore the local ionization barrier for a given ionic cell is assumed to be located at half distance to the next neighbor (${\bf r}_{\rm bar}$). The shift $\Delta_{\rm env}$ is then given by the electron potential produced from all charges (including the residual ion) at the point of the barrier (${\bf r}_{\rm bar}$) minus the electron potential in the environmental field alone (residual ion removed) sampled at the position of the ion. Ionization is tested if the condition $\hbar\omega>I_p^*$ is fulfilled. Cross sections for ionization steps that are energetically forbidden in free atomic Ar are approximated by the values for the highest possible atomic ionization step scaled with the remaining number of electrons in the tested atomic level.
For electron removal from the 3s level we assume instantaneous relaxation of the residual ion to the ground state. However, because of a much larger cross section, 3p photoionization is by far the dominant process in our case.

If an ionization event occurs, the atomic charge state is incremented and an electron is born at the position of the residual ion. The electron momentum is determined from energy conservation and initialized with random orientation. Active particles, i.e., all plasma electrons and ions, are propagated classically under the influence of the mutual Coulomb interaction and the laser field using a standard Verlet-algorithm. The coupled equations of motion read
\begin{equation}
m_i\ddot{\bf r}_i=q_ie{\bf E}_{\rm las}-\nabla_{{\bf r}_i}\sum_{i\neq j}V_{ij},
\end{equation}
where $m_i$, $q_ie$, and ${\bf r}_i$ are the mass, charge, and position of the $i$-th particle and ${\bf E}_{\rm las}={\bf e}_z E_0 f(t)\cos(\omega t)$ is the electric field of a linearly polarized laser pulse in dipole approximation with peak amplitude \mbox{$E_0=\sqrt{{2I_0}/{c\,\varepsilon_0}}$}. Here $c$ and $\varepsilon_0$ denote the vacuum speed of light and the vacuum permittivity. The pairwise Coulomb interaction $V_{ij}$ is described with a pseudopotential of the form
\begin{equation}
V_{ij}(r_{ij},q_1,q_2)=\frac{e^2}{4\pi \varepsilon_0}\frac{q_i q_j}{r_{ij}} \,{\rm erf}\left(\frac{r_{ij}}{s}\right)
\end{equation}
with the elementary charge $e$, the interparticle distance $r_{ij}$, their charge states $q_i$ and $q_j$, and a numerical smoothing parameter $s$. The latter regularizes the Coulomb interaction and offers a simple route to avoid classical recombination of electrons below the lowest possible quantum mechanical energy level. We use a constant value of $s$ and choose it to be the minimal value, such that the binding energy of an electron to a $q$-fold charged ion is always lower than the $q$-th ionization potential. For the particular case of argon this yields $s=1.128\,{\rm \AA}$~\cite{CommentSmooth}. We like to note that unphysical recombination occuring for lower values of the smoothing parameter can lead to artificial heating.

The above described MD scheme includes the full classical dynamics of plasma electrons and ions in the cluster and resolves thermalization and IBS heating in the nanoplasma. To analyze their impacts on the electron emission we will compare the MD simulations to results obtained from the simpler multistep model~\cite{BosPRL08}. The latter neglects any nanoplasma effects and assumes that cluster ionization proceeds as a series of instantaneous electron emission events due to direct photoemission from the developing cluster Coulomb field. Ionization is treated in the same way (via Monte-Carlo sampling) as in the MD code, but ionization events are accepted only if the single particle energy of the released electron is positive. The electrons are immediately removed from the simulation by neglecting any further energy exchange (instantaneous escape). In the final spectrum they occur with the single particle energy
\begin{equation}
E_{\rm sp} = \hbar \omega - I_{p}^{\alpha,i} - \frac{e^2}{4 \pi  \varepsilon_0} \sum_{i \neq j} \frac{q_j}{r_{ij}}
\label{eq_multistep}
\end{equation}
at the instant of birth (electron removal from level $\alpha$ of the $i$-th ion). The last term in Eq.\,(\ref{eq_multistep}) describes the Coulomb downshift due to previously generated ions with charge states $q_j$ at positions ${\bf r}_j$. For brevity, the resulting spectra are henceforwards termed Monte-Carlo results (MC) as opposed to the fully dynamical MD approach.

\section{Results and Discussion}
\label{sec_results}
In the following we present numerical results for the excitation of Ar$_{147}$ by 30\,fs pulses at \mbox{$\lambda=32\,{\rm nm}$}, i.e., for parameters similar to the experiments of Ref.\,\cite{BosPRL08}. To obtain sufficient statistics, ensemble averaging over 10$^{4}$ independent simulations is performed. Hence, all observables reflect statistical mean values.

\subsection{Phases of the Ionization Dynamics}
\begin{figure}[b]
     \centering \resizebox{1\columnwidth}{!} {\includegraphics{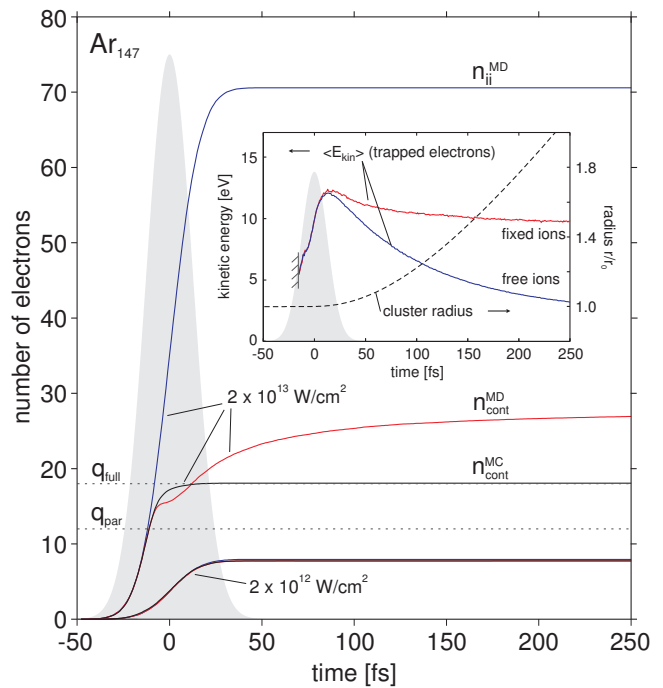}}
    \caption{(color online) Calculated ionization dynamics of Ar$_{147}$ for excitation with 30\,fs laser pulses at $\lambda=32\,{\rm nm}$ for two laser intensities (as indicated). The evolutions of total inner ionization $n_{\rm ii}$ and the number of continuum electrons $n_{\rm cont}$ are given for MD and MC simulations. For the lower intensity case all lines lie on top of each other. The shaded area depicts the laser intensity envelope. Dotted horizontal lines indicate ionization levels for partial ($q_{\rm par}$) and full ($q_{\rm full}$) frustration (see text). The inset displays the evolution of the average kinetic energy $\langle E_{\rm kin} \rangle$ of trapped electrons (negative single particle energy) from MD runs for $I=2 \times 10^{13} {\rm W/cm^2}$ with fixed and free ionic structure, respectively.}
    \label{fig_td_overview}
\end{figure}

To provide a basis for the forthcoming discussions, we begin with the analysis of characteristic stages of the interaction process. Fig.~\ref{fig_td_overview} displays the evolution of selected observables as predicted by MC and MD simulations for two representative laser intensities.
For MD both the number of innerionized electrons $n_{\rm ii}$ (equal to the total number of photoionization events) as well as the number of continuum electrons $n_{\rm cont}$ (positive single particle energy, equal to outer ionization) are displayed. For MC only the latter is given and serves as a reference for cluster charging when nanoplasma effects are suppressed.

For the lower intensity of \mbox{$I_0=2\times 10^{12}\rm\, W/cm^2$}, outer ionization evolves identical for both the MD and the MC approach. Electrons are excited directly to the continuum and practically no active electrons remain bound. This supports the claim of direct quasi-instantaneous electron emission without notable collision induced electron trapping. Hence, outer ionization is nearly proportional to the integrated pulse intensity and images the number of photons absorbed via photoionization. The dynamics can be fully described within the multistep picture.

The ionization behavior changes substantially for the higher intensity of \mbox{$I_0=2\times 10^{13}\rm\, W/cm^2$}. While MC and MD results still match in the rising edge of the pulse, the predictions begin to deviate at about \mbox{$t=-15\,\rm fs$}. From this point on, inner ionization starts to exceed outer ionization (MD). This reflects the onset of space charge trapping, where direct photoemission becomes partially frustrated, i.e., nanoplasma generation begins.

For a homogenous charge distribution in the cluster (jellium approximation) the trapping potential reads
\begin{equation}
\renewcommand{\arraystretch}{1.25}
V_{\rm jell}(q,r)=-\frac{q\,e^2}{4\pi\varepsilon_0}\left \{
\begin{array}{c l}
\frac{3R^2-r^2}{2R^3} & r \leq R\\
\frac{1}{r}  & r>R
\end{array}
\right.,
\label{eq_jellium}
\end{equation}
where $q$ is the total cluster charge (outer ionization prior to the considered ionization event) and \mbox{$R=r_s N^{1/3}$} is the cluster radius determined from the atomic Wigner-Seitz radius (\mbox{$r_s^{\rm Ar}=2.21 \AA$}). Partial frustration of direct photoemission begins in the cluster center if \mbox{$V_{\rm jell}(q_{\rm par},r=0)+\hbar\omega-I_p<0$}
, i.e., as soon as electrons are born with negative single particle energy. Considering 3p photoionization of a neutral Ar atom in the cluster, partial frustration sets in for a cluster charge state \mbox{$q_{\rm par}=12$}, in good agreement with the data in Fig.~\ref{fig_td_overview}. Subsequent ionization steps from regions where direct emission remains operational increase the depth of the space charge potential and eventually induce full frustration of direct photoemission if
\mbox{$V_{\rm jell}(q_{\rm full},r=R)+\hbar\omega-I_p<0$}.
Note that the thresholds for partial and full frustration are related by $q_{\rm full}/q_{\rm par}=3/2$ within this electrostatic approximation.
In our case, full frustration is reached for a cluster charge of \mbox{$q_{\rm full}=18$}, which reflects the upper limit for multistep ionization.

Whereas the MC data shows smooth convergence of outer ionization to the full frustration limit \mbox{$q_{\rm full}$}, a transient saturation at a slightly lower value is found near \mbox{$t=-5\,$fs} in the corresponding MD result. This indicates a small enhancement of the trapping field by electrons that are produced in the center of the cluster, i.e., with insufficient energy to escape. More importantly, a second long-living emission feature emerges around \mbox{$t\approx 0\,\rm fs$} and outer ionization finally increases to values well beyond $q_{\rm full}$. This additional emission persists after the laser pulse and can be attributed to thermal electron evaporation from the nanoplasma.

For a closer analysis of this evaporative emission we examine the evolution of the nanoplasma in more detail. The total number of trapped electrons is given by \mbox{$n_{\rm trapped}=n_{\rm ii}-n_{\rm cont}$}, cf. Fig.~\ref{fig_td_overview}. Note that the nanoplasma population strongly exceeds the number of directly emitted electrons for the higher intensity case. The maximum evaporation rate of thermal electrons occurs near \mbox{$t=15\,\rm fs$}, i.e, even before the maximum number of trapped electrons is reached. The reason for that is the competition between population and temperature in the nanoplasma. The evolution of the average kinetic energy of trapped electrons shows a heating phase up to about \mbox{$t=15\,\rm fs$} and rapid cooling afterwards, see inset of Fig.~\ref{fig_td_overview}. The peak of this average thermal energy well coincides with the maximum evaporation rate, underlining the strong temperature dependence of evaporative emission. Two processes contribute to the nanoplasma cooling: (i) the evaporation itself and (ii) cluster expansion. The latter is driven by both Coulomb forces from cluster charging as well as thermal pressure and results in very efficient expansion cooling of trapped electrons. A comparison of runs with free vs. frozen ions shows that expansion cooling is the dominant process, see inset of Fig.~\ref{fig_td_overview}.
The rapid decay of the evaporation rate after the laser pulse is therefore mainly due to expansion cooling of nanoplasma electrons. Note that thermal evaporation is almost completed around $t=150\,\rm fs$.

The above analysis shows that the XUV induced electron emission proceeds in two fundamentally different stages, i.e., first via multistep ionization and subsequently by evaporative emission from the nanoplasma. For reaching the evaporative regime, the cluster ionization has to be sufficiently high to frustrate direct photoemission, i.e., to generate the nanoplasma.

\subsection{Electron Emission Spectra}
\label{sec_elspectra}
\begin{figure*}[t]
     \centering \resizebox{2\columnwidth}{!} {\includegraphics{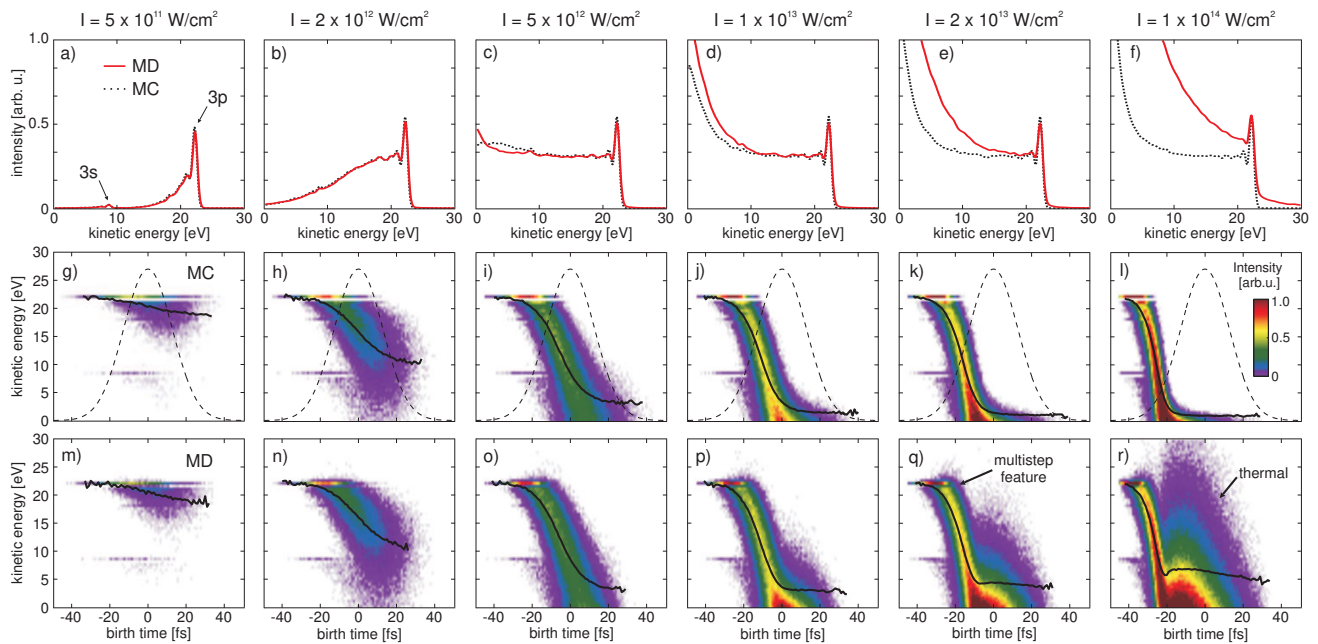}}
    \caption{(color online) Intensity dependent electron spectra from Ar$_{\rm 147}$ (from left to right). The upper panels (a-f) compare final energy distributions calculated from MC (dotted) and MD (solid) simulations. Middle and lower panels display an energy vs. birth time analysis. Therefore the final energy distributions (color-coded) and mean energies (solid curve) of electrons are given as function of their release time from the mother ion. Hereby the middle row [panels (g)-(l)] displays the data from MC runs whereas the lower row reflects corresponding MD results. Characteristic signatures from multistep ionization and thermal electron evaporation are indicated in panels (q) and (r). The laser intensity envelope is schematically shown (dashed).}
    \label{fig_electron_spectra}
\end{figure*}

In the next step we analyze characteristic signatures of the above described ionization stages in the electron emission spectra. Fig.~\ref{fig_electron_spectra} displays the evolution of the final energy distributions of emitted electrons as function of laser intensity (from left to right, upper panels) for MC and MD simulations.
An energy vs. birth time analysis (middle and lower panels) allows assignment of particular features to different electron release times.

For intensities up to about $I=5\times 10^{12}\,{\rm W/cm^2}$, the respective MC and MD spectra are almost identical, see Figs.~\ref{fig_electron_spectra}(a)-\ref{fig_electron_spectra}(c), and reflect the typical behavior of multistep ionization. The multistep process produces a shoulder at the low energy side of the atomic photoline, which broadens and develops into a plateau as intensity increases. Note that the main contribution is due to the emission of 3p-electrons, cf. Fig.~\ref{fig_electron_spectra}(a). Starting at the energy of the atomic photoline \mbox{$E=\hbar\omega-I_p$} (first electron), electron energies are continuously downshifted for each ionization step due to the cluster space charge, as reflected by the signatures in the energy vs birth time evolutions, see Figs.~\ref{fig_electron_spectra}(g)-\ref{fig_electron_spectra}(i) and Figs.~\ref{fig_electron_spectra}(m)-\ref{fig_electron_spectra}(o). Note that the latter are practically equal for MC and MD (middle and lower panels) up to \mbox{$I=5\times10^{12}\rm W/cm^2$}. The emergence of the plateau in the energy spectrum can be understood by recalling the trapping potential in jellium approximation given in Eq.~(\ref{eq_jellium}). According to this, emitted electrons show an average energy downshift with respect to the atomic photoline of
\begin{equation}
\langle \Delta E(q) \rangle =-\frac{e^2}{4\pi\varepsilon_0}\frac{6q}{5R},
\end{equation}
which scales linearly with charge state. As the average energy distribution for a given ionization step exhibits a finite width (except for the first electron), the contributions add up to a plateau-like spectrum eventually.

For intermediate laser intensity [Fig.~\ref{fig_electron_spectra}(d)] an additional feature appears at low electron energies on top of the plateau in the MC and the MD spectra. It can be traced back to selective photoemission resulting from partial frustration, where only electrons from outer regions of the cluster can be emitted directly. The resulting reduced energy downshift per ionization step increases the signal at low energies. Full frustration of direct photoemission at higher intensities is reflected in the rapid saturation of MC spectra, see Figs.~\ref{fig_electron_spectra}(d)-\ref{fig_electron_spectra}(f). Though the MD spectra begin to substantially exceed the MC results in this intensity range due to thermal evaporation, the characteristic energy downshift with birth-time (multistep feature) persists in early stages for all scenarios, see Figs.~\ref{fig_electron_spectra}(j)-\ref{fig_electron_spectra}(l) and Figs.~\ref{fig_electron_spectra}(p)-\ref{fig_electron_spectra}(r).

The departure from pure multistep behavior for \mbox{$I \gtrsim 10^{13}{\rm W/cm^2}$} is in agreement with experimental findings ~\cite{BosPRL08} and numerical results from kinetic Boltzmann simulations~\cite{Ziaja2009}. The roughly exponential additional contribution in the energy spectra due to electron evaporation exhibits a characteristic signature in the birth-time analysis. Comparison of the energy vs birth time plots of MD and MC runs reveals that thermal evaporation shows up as an additional broad feature in the MD results (see Figs.~\ref{fig_electron_spectra}(p)-Fig.~\ref{fig_electron_spectra}(r) as soon as full frustration is reached.
The onset of evaporation induces a turnover of the average electron energy vs birth-time (solid curve) from a decaying to an almost constant evolution. The latter effect is most pronounced in Figs.~\ref{fig_electron_spectra}(q) and \ref{fig_electron_spectra}(r). Interestingly, the average final energy of thermally emitted electrons is only weakly dependent on the laser intensity. Focussing on the time after the pulse peak, the average energy increases by less than a factor of two when going from \mbox{$2\times10^{13}$} to \mbox{$10^{14}\,\rm W/cm^2$}, cf. Figs.~\ref{fig_electron_spectra}(q)-\ref{fig_electron_spectra}(r). Though more intense pulses produce a higher number of thermal electrons, their average energy changes only weakly. This is a central result of our analysis and indicates a characteristic heating process. We come back to this aspect in Sec.~\ref{sec_ionization_heating}.

\begin{figure}[t]
    \centering \resizebox{\columnwidth}{!} {\includegraphics{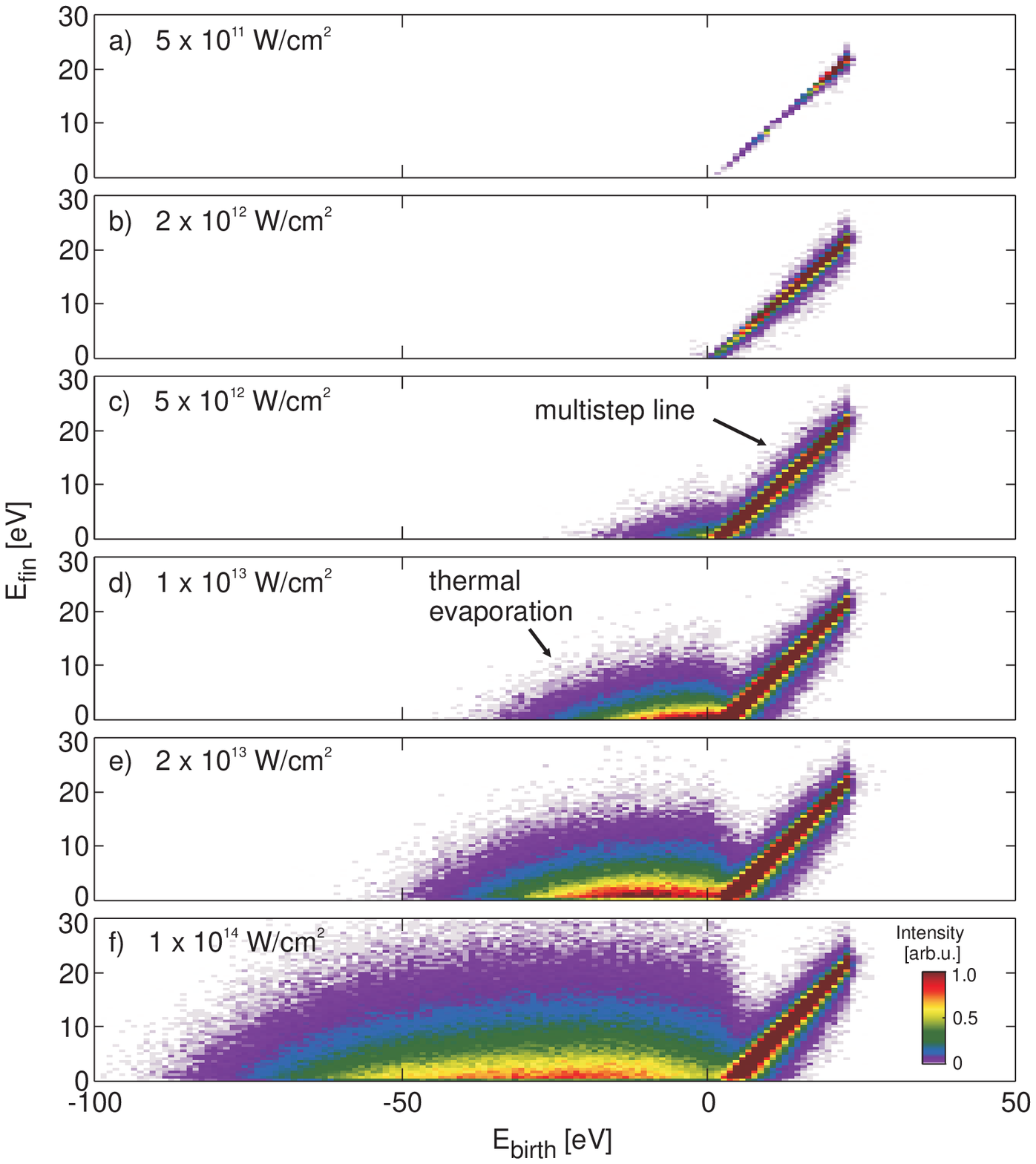}}
    \caption{(color online) Energy correlation analysis of emitted electrons for the MD simulations of Fig.~\ref{fig_electron_spectra}. The plots displays the final kinetic energy distributions ($E_{\rm fin}$) of electrons as function of their single particle energies at the instant of birth ($E_{\rm birth}$). Characteristic features from multistep ionization and thermal evaporation are indicated.}
    \label{fig03}
\end{figure}

Both the persistence of multistep ionization at high intensity (in early stages) and the emergence of evaporative emission are corroborated by an analysis of the correlation of single particle energies of electrons at the instant of birth ($E_{\rm birth}$) with their final energy after emission ($E_{\rm fin}$). Figure~\ref{fig03} contains a corresponding correlation plot for the MD runs of Fig.~\ref{fig_electron_spectra}.
Up to about \mbox{$I=5\times 10^{12}{\, \rm W/cm^2}$} the birth- and final energies are highly correlated (almost equal), forming a characteristic multistep line around $E_{\rm fin}=E_{\rm birth}\ge0$, cf. Figs.~\ref{fig03}(a)-\ref{fig03}(c). The latter is found in all runs, but its increasing width at higher laser intensity indicates the growing effect of energy exchange between multistep electrons during their escape. Emitted electrons with initial birth energies $E_{\rm birth}<0$ begin to appear for \mbox{$I\gtrsim5\times 10^{12}{\,\rm W/cm^2}$}. Their uncorrelated $E_{\rm fin}$ and $E_{\rm birth}$ values reflect the (random) energy exchange with other nanoplasma electrons required for evaporative emission. Besides being a marker for electron emission processes, the energy correlation analysis contains an image of the cluster potential during the interaction time with the laser pulse. The lower $E_{\rm birth}$ cutoff of the distribution is a measure for the depth of the trapping field in the cluster. With increasing intensity the minimum birth energies decrease and reach values of about $E_{\rm birth}^{\rm min}=-90\,{\rm eV}$ at $I=10^{14}{\rm W/cm^2}$, see Fig.~\ref{fig03}(f). The fact that the depth of the space charge potential becomes much larger than the photon energy reflects the growing contribution of temperature-driven charge separation to the cluster potential (thermal spill-out of nanoplasma electrons).

\subsection{Signatures of ionization heating}
\label{sec_ionization_heating}
\begin{figure}[t]
    \centering \resizebox{0.85 \columnwidth}{!} {\includegraphics{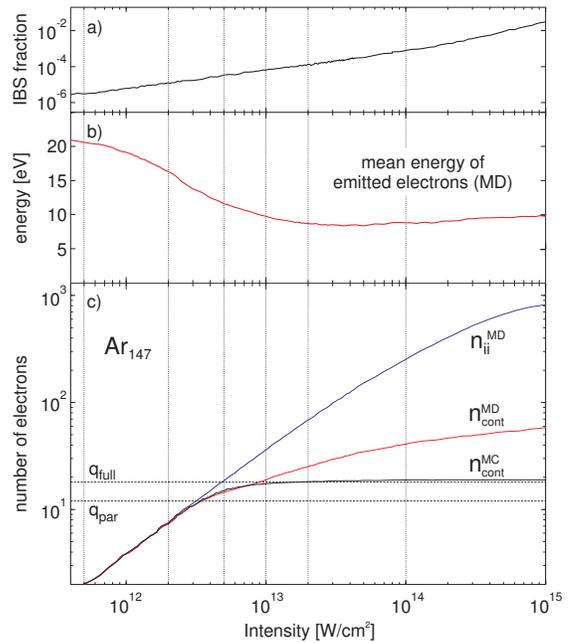}}
    \caption{(color online) Intensity dependent heating and ionization of Ar$_{\rm 147}$ in 30\,fs laser pulses at $\lambda=32\,{\rm nm}$. The upper and middle panel show the relative fraction of IBS heating to the total energy absorption and the average energy of emitted electrons, respectively. The lower panel displays the final values of total inner ionization ($n_{\rm ii}$) and the number of continuum (emitted) electrons ($n_{\rm cont}$) for MD and MC simulations. Dashed horizontal lines indicate ionization levels for partial ($q_{\rm par}$) and full ($q_{\rm full}$) frustration. Vertical dotted lines indicate laser intensities of the scenarios discussed in Figs.~\ref{fig_electron_spectra} and \ref{fig03}.}
    \label{fig04}
\end{figure}

We now return to a closer analysis of the interrelation of ionization and heating processes in the nanoplasma.
Figure~\ref{fig04}(c) displays a systematic comparison of final inner ($n_{\rm ii}^{\rm MD}$) and outer cluster ionization ($n_{\rm cont}^{\rm MD}$) from MD runs with outer ionization from MC simulations ($n_{\rm cont}^{\rm MC}$) as function of intensity.
The similar evolutions of outer ionization from MC and MD reflect pure multistep ionization up to about \mbox{$I_0=3\times10^{12}\,\rm W/cm^2$}. Note that the partial frustration threshold is approximately reached for this intensity. Beyond that intensity the nanoplasma formation begins (deviation of inner and outer ionization in the MD results), but multistep ionization prevails until full frustration sets in near $I_0=10^{13}\,\rm W/cm^2$. Up to this point the mean energy of emitted electrons (MD), see Fig.~\ref{fig04}(b), decreases steadily with laser intensity and approaches a value of 50\% of the energy of the atomic 3p photoline, reflecting roughly the center of mass of the multistep plateau in the energy spectra, cf. Fig.~\ref{fig_electron_spectra}. For higher intensity thermal evaporation [difference between outer ionization of MD and MC in Fig.~\ref{fig04}(c)] becomes substantial and exceeds the fraction of multistep electrons near \mbox{$I_0=8\times10^{14}\,\rm W/cm^2$}. The degree of inner ionization grows steadily with pulse power and reaches values up to $\langle q_{\rm ii}\rangle=5.5$ (per atom) in the displayed intensity range. This results in a very dense nanoplasma and underlines the importance of ionization threshold lowering in the cluster.

However, even if a dense nanoplasma is formed, the relative impact of IBS heating to the total energy absorption (MD) remains negligible ($\lesssim 1\%$ for the displayed intensity range), see Fig.~\ref{fig04}(a). This is in strong contrast to the VUV domain, where nanoplasma heating via IBS is substantial and typically even the dominating absorption channel \cite{SiePRL04,GeoPRA07}. The vanishingly small contribution of IBS heating in the XUV case implies that the excess energy from inner photoionization acts as the main source of thermal energy in the nanoplasma.
We thus suggest to term this energy absorption process {\it ionization heating}. A direct consequence of the latter is that the nanoplasma temperature cannot be pushed unlimitedly by just increasing the laser intensity.
Because of the temperature dependence of electron evaporation, heating processes leave clear traces in the intensity dependent energy of emitted electrons, see Fig.~\ref{fig04}(b). In fact, the presence of ionization heating leads to a very unusual evolution of the mean electron energy, showing an almost constant value from the onset of nanoplasma generation up to high intensities. The excitation with intense short-wavelength pulses thus offers a route to produce transient nanoplasmas with well controlled density (tunable by the intensity) at nearly constant temperature (adjustable with the photon energy).

\section{Summary and Conclusions} \label{sec_conclusions}
The stages and intensity dependence of charging and heating in rage-gas clusters under intense XUV laser pulses have been analyzed by comparison of molecular dynamics simulations with statistical Monte-Carlo results. We find that the electron emission proceeds in two sequential phases: first via direct photoemission and then via thermal electron evaporation. The applicability of the concept of multistep ionization up to high laser intensities - at least in the rising edge of the pulse - has been demonstrated and estimates for the thresholds for partial and full frustration of direct photoemission have been given based on a simplified jellium picture. Thermal electron evaporation occurs as soon as a nanoplasma is formed. The latter requires frustration of direct photoemission and is thus delayed by a certain number of ionization steps. The major heating mechanism of the nanoplasma is a process we term {\em ionization heating}, i.e., inner photoionization of localized electrons, while IBS absorption of plasma electrons is negligible.

Both multistep ionization and ionization heating are generic mechanisms that are of relevance for a wide class of laser-particle interactions in the wavelength range from the XUV up to x-ray domain. For example, intense laser irradiation required for single-shot x-ray diffractive imaging of finite samples produces a deep trapping potential due to multistep ionization, which is crucial for secondary ionization processes. A sufficiently strong trapping field  triggers field ionization cascades, which can lead to charge migration processes~\cite{GnoPRA09} and substantial depletion of the number of bound electrons. A suppression of the latter is of importance for retrieving useful diffraction pattern from the sample~\cite{NakPRA09}. Another direction is the controlled generation of small plasma targets, whose density and temperature could in principle be tuned over a wide range. While the intensity mainly controls the density (if the frustration threshold is reached), the photon energy can be varied to control the temperature produced via ionization heating. We thus expect that the mechanisms of multistep ionization and ionization heating will not only be of fundamental importance for laser-cluster science, but have far reaching implications for the rapidly growing field of intense short-wavelength laser-matter interactions in general.

\begin{acknowledgements}
We like to thank Christian Gnodtke and Ulf Saalmann for inspiring discussions with regard to the interpretation of the birth time analysis. Financial support by the Deutsche Forschungsgemeinschaft within the SFB 652/2 and computer time provided by the HLRN Computing Center are gratefully acknowledged.
\end{acknowledgements}

\bibliographystyle{apsrev}

\begin{thebibliography}{10}
\bibitem{CommentSmooth}
The same smoothing parameter for electron-electron, ion-ion, and electron-ion interactions is used to avoid force discontinuities for surrounding particles during an ionization event. A natural but often acceptable drawback of this treatment is that the effect of energetic collisions at low impact parameters is underestimated.
\end{thebibliography}

\end{document}